\documentclass[journal]{IEEEtran}
\ifCLASSOPTIONcompsoc
\usepackage[nocompress]{cite}

\else
\usepackage{cite}
\fi

\ifCLASSINFOpdf
   \usepackage[pdftex]{graphicx}
\else
\fi

\newcommand{\sapphire}{sapphire~}
\newcommand{\alscn}{Al\textsubscript{0.72}Sc\textsubscript{0.28}N~}
\newcommand{\alscnNospace}{Al\textsubscript{0.72}Sc\textsubscript{0.28}N}
\newcommand{\alscnX}{Al\textsubscript{1-x}Sc\textsubscript{x}N~}
\newcommand{\alscnXNospace}{Al\textsubscript{1-x}Sc\textsubscript{x}N}

\begin{document}

%
\title{Ultrathin \alscnX for low-voltage driven ferroelectric-based devices}
%
%
%

\author{Georg Schönweger,~\IEEEmembership{Student Member,~IEEE}, Md Redwanul Islam,~\IEEEmembership{Student Member,~IEEE}, Niklas Wolff, Adrian Petraru, Lorenz Kienle, Hermann Kohlstedt, Simon Fichtner,~\IEEEmembership{Member,~IEEE}%
}

%
%

\markboth{Journal of \LaTeX\ Class Files,~Vol.~14, No.~8, August~2015}%
{Shell \MakeLowercase{\textit{et al.}}: Bare Demo of IEEEtran.cls for IEEE Journals}
%



\maketitle

\begin{abstract}
Thickness scaling of ferroelectricity in \alscnX is a determining factor for its potential application in neuromorphic computing and memory devices. In this letter, we report on ultrathin (10 nm) \alscn films that are ferroelectrically switchable at room temperature. All-epitaxial \alscnNospace/Pt heterostructures are grown by magnetron sputtering onto GaN/\sapphire substrates followed by an \textit{in situ} Pt capping approach to avoid oxidation of the \alscn film surface. Structural characterization by X-ray diffraction and transmission electron microscopy reveals the established epitaxy. The thus obtained high-quality interfaces in combination with the \textit{in situ} capping is expected to facilitate ferroelectric switching of \alscnX in the ultrathin regime. The analysis of the relative permittivity and coercive field dependence on the \alscn film thicknesses in the range of 100 nm down to 10 nm indicates only moderate scaling effects, suggesting that the critical thickness for ferroelectricity is not yet approached. Furthermore, the deposited layer stack demonstrates the possibility of including ultrathin ferroelectric \alscnX into all-epitaxial GaN-based devices using sputter deposition techniques. Thus, our work highlights the integration and scaling potential of all-epitaxial ultrathin \alscnX offering high storage density paired with low voltage operation desired for state of the art ferroelectric memory devices.
\end{abstract}

\begin{IEEEkeywords}
aluminum-scandium-nitride (\alscnXNospace), ultrathin, ferroelectric, gallium nitride,  epitaxial growth
\end{IEEEkeywords}

%
\IEEEpeerreviewmaketitle

\section{Introduction} 
\IEEEPARstart{T}{he} high coercive field ($E_c$) and the high, stable remanent polarization separates the ferroelectric properties recently discovered in materials with wurtzite-type structure from classical ferroelectrics\cite{Fichtner2019,Ferri_2021,Hayden2021,Wang2021b}. This raises hopes for particularly good scalability of wurtzite-type based ferroelectric devices. In addition, the CMOS compatibility and the well established industrial deposition process make \alscnX thin films highly attractive for building novel neuromorphic computing and memory devices such as ferroelectric field-effect transistors (FeFET) and ferroelectric tunnel junctions (FTJ)\cite{Mulaosmanovic2020,Schenk2020,Tsymbal_2006,Kohlstedt2005,liu2020postcmos}. Furthermore, it is expected that wurtzite-type ferroelectrics such as \alscnX introduce ferroelectricity into III-N technology, resulting in a straightforward approach to realize embedded memory. Recently, ferroelectric all-epitaxial all-wurtzite type \alscnXNospace/GaN heterostructures were demonstrated\cite{Wang2021a,Schoenweger2022}. However, a very low film thickness of the ferroelectric layer is needed for following the general trend of miniaturization and increasing storage density in all of the aforementioned devices. In this context, \alscnX offers high scalability due to its high $E_c$, making it possible to tailor the film thickness to the ultrathin regime to achieve reasonable memory windows and low operating voltages\cite{Mulaosmanovic2019}. Furthermore, in terms of device design, ultrathin ferroelectric films are a prerequisite for building FTJs\cite{Tsymbal_2006}. However, reducing the thickness to the ultrathin regime ($<$ 30 nm) is often accompanied with a material-specific diminution of the remanent polarization or results in a total loss of ferroelectricity\cite{Qiao2021}. Up to now, only a small number of studies on the thickness scaling properties of ferroelectric \alscnX were conducted, revealing ferroelectricity for $\approx$20 nm film thickness at room temperature as well as for $\approx$10 nm film thickness at elevated temperatures (373 K)\cite{Fichtner2020a,Wang2020,Mizutani2021}. Here we present our recent results on 10 nm thick epitaxial \alscn grown on epitaxial Pt/GaN templates using sputter deposition. Typical butterfly-shaped capacitance-voltage ($C-U$) loops were recorded at room temperature, demonstrating clearly distinguishable ferroelectric switching. Thus, the general feasibility of  ultrathin \alscnX for future device integration and low voltage operation is demonstrated.

\section{Device fabrication and characterization methods}
All films were grown by sputter deposition in an Oerlikon (now Evatec) MSQ 200 multisource system on top of commercially available GaN/\sapphire substrates. A 12 nm thick epitaxial Pt layer (bottom electrode) was DC-sputtered at 500 °C, 600 W, and 50 sccm Kr flow. XRD measurements were performed using monochromatic K\textsubscript{$\alpha$} radiation in a Seifert XRD 3000 PTS system ($\theta$-2$\theta$ scan) as well as in a  Rigaku SmartLab diffractometer (Phi-scan). The \alscn thin films were grown by pulsed-DC co-sputtering at 450 °C, details about the process can be found elsewhere\cite{Fichtner2015,Fichtner2017}. The 10 nm thick \alscn layer was capped \textit{in situ} with 100 nm thick Pt. Square top-electrodes were structured via lithography and ion beam etching (IBE, Oxford Instruments Ionfab 300). Capacitance measurements were performed on a Hewlett Packard 4284A Precision LCR meter. A cross-section specimen of the \textit{in situ} capped 10 nm thick \alscn sample has been prepared using a standard focused ion beam (FIB) procedure and analyzed using high-resolution transmission electron microscopy (HRTEM, Tecnai F30, operated at 300 kV, field emission gun) and Fast Fourier Transform (FFT) patterns.

\section{Results and Discussions}
\begin{figure}[h!]
\centering
\includegraphics[]{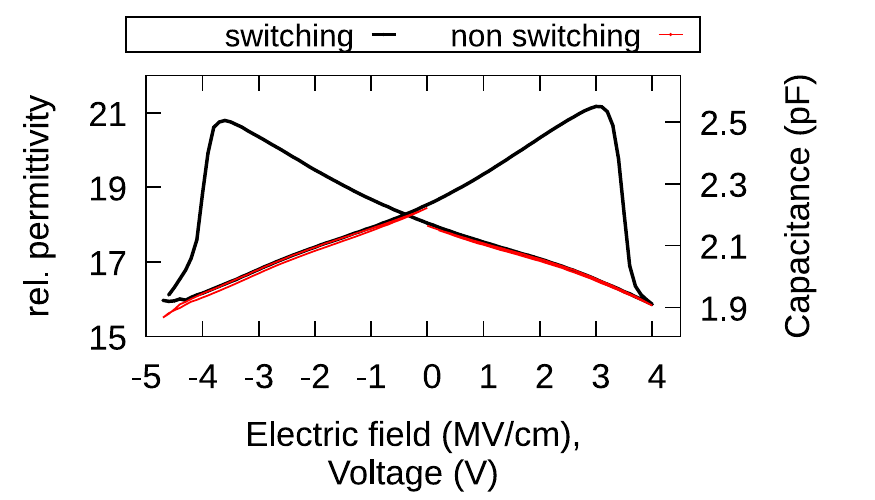}
\caption{Relative permittivity and capacitance as a function of the applied voltage and electric field for Pt(100 nm)/\alscnNospace(10 nm)/Pt(12 nm) capacitors deposited on a GaN/\sapphire template. The non switching half-loops (red) were recorded by previously switching the capacitor to the respective polarity. Measured on 10 x 10 \textmu m² pads with a small signal of 1 V\textsubscript{pp} at 100< kHz swept in 0.1 V steps with a delay of 300 ms per step.}
\label{fig:c-v-10nm}
\end{figure}

\begin{figure}[h!]
\centering
\includegraphics[]{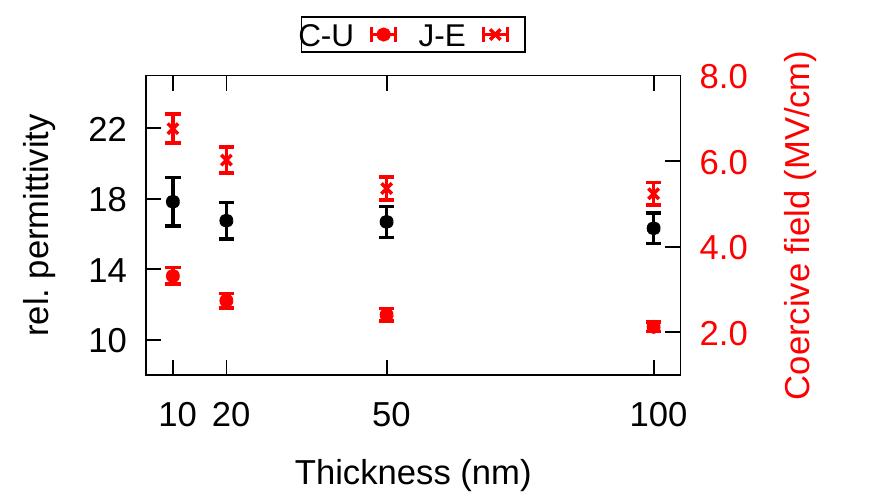}
\caption{$\varepsilon_{r}$ and $E_c$ in dependence of \alscn film thickness. $\varepsilon_{r}$ was measured at 100 kHz, 100 mV at 0 V bias. $E_c$ was determined via measuring $C-U$ loops keeping the time for a full sweep constant for the various film thicknesses as well as via $J-E$ using a sinus signal at 80 kHz. The error bars were calculated using an estimated capacitor side-length error of 0.4 \textmu{}m, an estimated thickness error of 5 \% and an estimated error for determination of the $E_c$. Additionally for $E_c$ determination via $C-U$ loops the step width of the respective voltage sweep was included in the error calculation.}
\label{fig:e_r-E_c-vs-thickness}
\end{figure}

The voltage dependence of the relative permittivity ($\varepsilon_{r}$) is depicted in Fig. \ref{fig:c-v-10nm}. The butterfly-shaped curve (black) demonstrates distinguishable ferroelectric switching of the 10 nm thick \alscn film.
Additionally non switching half-loops (red) were recorded by measuring twice in the same direction. No hysteretic behaviour is observed for the non-switching half-loops, thus giving additional evidence that the butterfly-shaped hysteresis for the switching loop originates from ferroelectric switching. Previously, the coercive field in \alscnX thin films was observed to be almost independent of film thickness down to 27 nm\cite{Fichtner2020a}. Similarly, the $E_c$ fields determined via $C$-$U$ loops in this study, as depicted in Fig. \ref{fig:e_r-E_c-vs-thickness}, is not changing considerably when decreasing the film thickness from 100 nm to 50 nm, while at 20 nm and escpecially at 10 nm film thickness, a moderate increase of the coercive field by about 20\% is observed.

This is in strong contrast to classical ferroelectrics, e.g. the perovskites (PZT, BTO, etc.), were a pronounced increase of $E_c$ with decreasing film thickness is observed\cite{Pertsev2003,Qiao2021}. Furthermore, the often reported degradation of ferroelectric properties due to a non-switching interfacial layer (dead layer\cite{Zhou1997}) in ultrathin films is typically accompanied by a decrease of $\varepsilon_{r}$\cite{Pertsev2003,Stengel2006}. Thus,  ferroelectricity in 10 nm thick \alscnX appears to be not yet influenced considerably by such interfacial effects. The non-scaling of $E_c$ and $\varepsilon_{r}$ with thickness leads to our conclusion that a critical thickness in \alscnX is not yet approached at 10 nm and stable ferroelectricity can be expected also for films $<$ 10 nm thickness.

\begin{figure}[h!]
\centering
\includegraphics[]{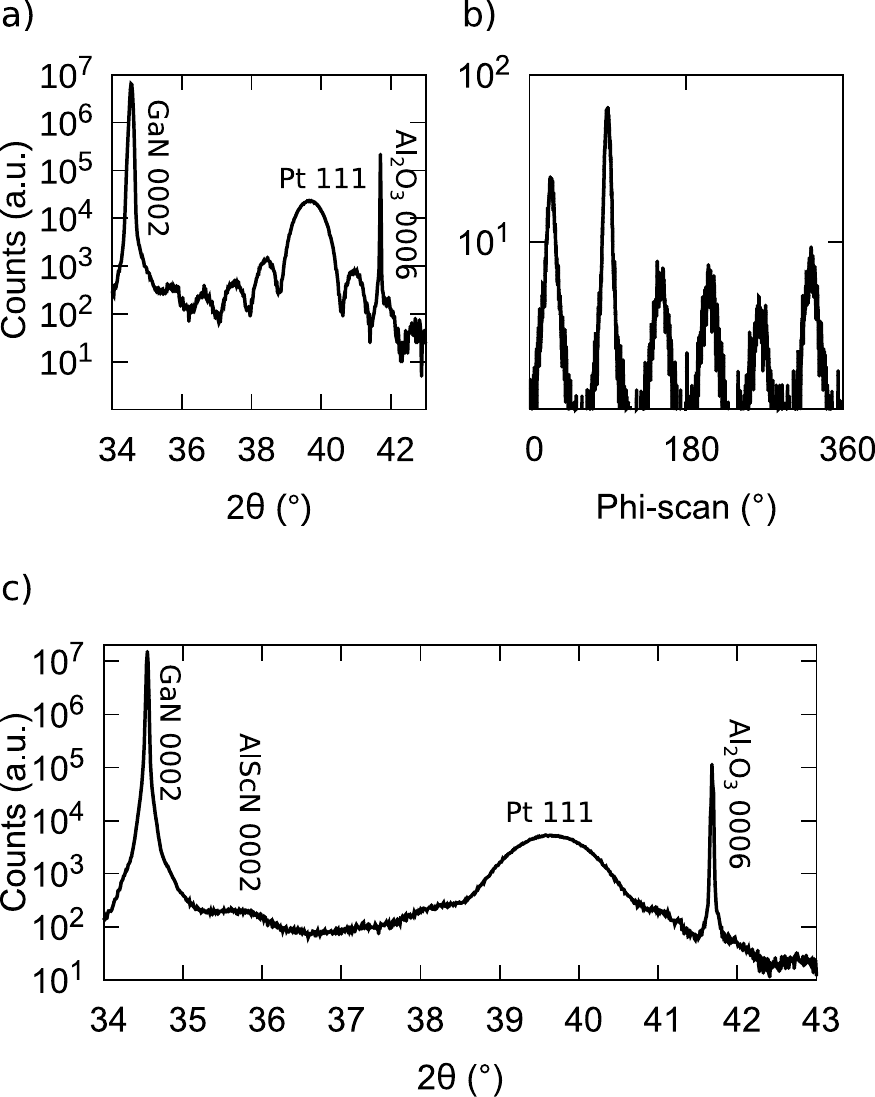}
\caption{a) $\theta$-2$\theta$ scan and b) smoothed phi-scan for the non-specular Pt 131 reflection (right) of 12 nm Pt on GaN/\sapphire. c) $\theta$-2$\theta$ of the film shown in a) after deposition of 10 nm \alscn followed by 5 nm in-situ capping with Pt. } 
\label{fig:t2t-epi-Pt} 
\end{figure}

The $\theta$-2$\theta$ scan depicted in Fig. \ref{fig:t2t-epi-Pt} (left) reveals the expected out-of-plane 111 orientation for Pt and 0002 orientation for GaN. Laue oscillations, which are sensitive to crystalline disorder, are appearing for the Pt 111 reflection\cite{Castel1998}.
The thus expected high crystallinity is confirmed by rocking curve (RC) measurements (not shown here) of the 111 Pt reflection; the full width at half maximum (FWHM) of 0.1 ° is indicative of a strain and defect-poor crystalline phase with high out-of-plane orientation. Furthermore, the distinct Laue oscillations allow to determine the thickness ($t$) of the ordered crystal volume\cite{Castel1998}.
\begin{equation} \label{eq:laue-thickness}
t = \frac{3 \lambda}{2 \sin{\theta_{1}} - \sin{\theta_{-1}}}
\end{equation}

Using equation \ref{eq:laue-thickness}, with $\theta_{1}$ and $\theta_{-1}$ corresponding to the maximum of the peaks right and left from the center Pt 111 reflection, an average film thickness of 11 nm is determined. 

\begin{figure}[h!]
\centering
\includegraphics[]{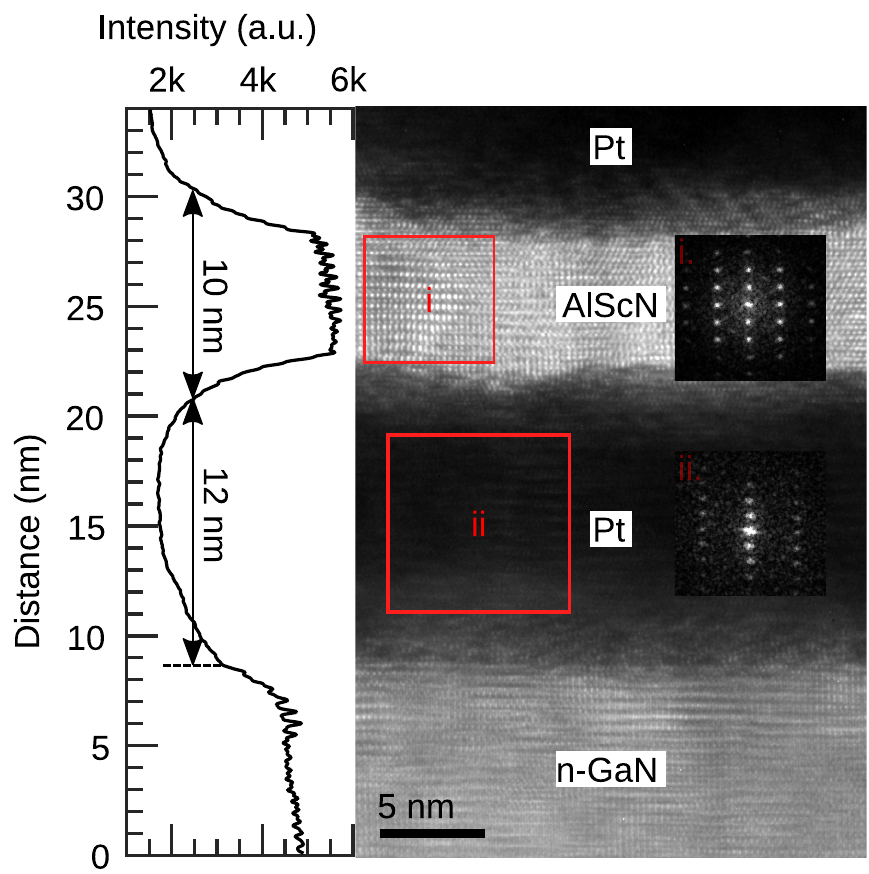}
\caption{HRTEM image (right) and corresponding averaged thickness profile (left) of the Pt(100 nm)/\alscnNospace(10 nm)/Pt(12 nm)/GaN/\sapphire heterostructure. In the inset, the FFT pattern for \alscn (i) and Pt (ii) is depicted.} 
\label{fig:hrtem} 
\end{figure}

The in-plane ordering was resolved by recording a phi-scan of the non-specular 131 Pt reflection, depicted in Fig. \ref{fig:t2t-epi-Pt} (right). A 6-fold symmetry is visible, evidencing epitaxial (in-plane ordered) growth.

The intensity profile across the HRTEM micrograph is shown in Fig. \ref{fig:hrtem} and film thicknesses of 10 nm for \alscn and 12 nm for Pt are determined on local average. The latter is in excellent agreement with the Pt thickness calculated using equation \ref{eq:laue-thickness}. The FFT pattern clearly reveals an epitaxial growth for both layers. The epitaxial growth in combination with the \textit{in situ} capping of \alscnX to avoid oxidation is expected to have a significant impact on the ferroelectric switching behaviour, especially for ultrathin films. The blurred interfaces present in the HRTEM image are attributed to the finite roughness of the heterostructure.

\newpage
\section{Conclusion}
Distinguishable ferroelectric switching of 10 nm thick \alscn was demonstrated at room temperature. Largely constant coercive fields and constant $\varepsilon_{r}$ for films ranging from 100 nm down to 10 nm thickness were confirmed. Thus, no indications of instabilities of the ferroelectric phase or considerable contributions from dead layers are present. Therefore, it is expected that further potential remains to scale ferroelectric \alscnX to even lower thicknesses. These results highlight the general feasibility of using \alscnX as an active layer for integrated electronic devices with low operating voltage and high storage density.


%

\section*{Acknowledgment}
This work was supported by the project “ForMikro-SALSA” (grant no. 16ES1053) from the Federal Ministry of Education and Research (BMBF) and the Deutsche Forschungsgemeinschaft (DFG) under the scheme of the collaborative research centers (CRC) 1261 and 1461.

\ifCLASSOPTIONcaptionsoff
  \newpage
\fi




\bibliographystyle{IEEEtran}
\bibliography{IEEEabrv,literature-v2.bib}


%

%




\end{document}